\documentclass[prl,aps,twocolumn,nofootinbib,10pt, showpacs]{revtex4-1}
\usepackage{amssymb}
\usepackage{times}
\usepackage{graphicx}
\usepackage{subeqnarray}
\newcommand{\ket}[1]{|#1\rangle}
\newcommand{\bra}[1]{\langle #1|}

\begin{document}

\title{Topological Crystalline Insulator and Quantum Anomalous Hall States in IV-VI based 
       Monolayers and their Quantum Wells}

\author{Chengwang Niu}
\email{c.niu@fz-juelich.de}
\author{Patrick M. Buhl}
\author{Gustav Bihlmayer}
\author{Daniel Wortmann} 
\author{Stefan Bl\"{u}gel}
\author{Yuriy Mokrousov}
\affiliation{Peter Gr\"{u}nberg Institut and Institute for Advanced Simulation, 
             Forschungszentrum J\"{u}lich and JARA, 52425 J\"{u}lich, Germany}          

\begin{abstract}
Different from the two-dimensional (2D) topological insulator, the 2D topological crystalline 
insulator (TCI) phase disappears when the mirror symmetry is broken, e.g., upon placing on a 
substrate. Here, based on a new family of 2D TCIs $-$ SnTe and PbTe monolayers $-$ we 
theoretically predict the realization of the quantum anomalous Hall effect with Chern number 
$\mathcal C = 2$ even when the mirror symmetry is broken. Remarkably, we also demonstrate 
that the considered materials retain their large-gap topological properties in quantum well 
structures obtained by sandwiching the monolayers between NaCl layers. Our results demonstrate 
that the TCIs can serve as a seed for observing robust  topologically non-trivial phases. 
\end{abstract}

\maketitle
\date{\today}

The topological crystalline insulator (TCI) is a recently discovered new class of quantum 
states with insulating bulk energy gap and gapless edge or surface states~\cite{Fu}. Gapless 
edge or surface states of TCIs arise from the crystalline symmetry in contrast to that of 
the $\mathbb{Z}_2$  topological insulator (TI), where the reason for protection is  
time-reversal symmetry (TRS)~\cite{Fu,Hsieh,Hasan,Qi1}. The first proposal of a material class 
with TCIs was purely theoretically~\cite{Hsieh} and consequently TCIs were experimentally 
observed in three-dimensional (3D) IV-VI semiconductors SnTe~\cite{Tanaka}, 
Pb$_{1-x}$Sn$_x$Se~\cite{Dziawa}, and Pb$_{1-x}$Sn$_x$Te~\cite{Xu,Yan}. Later on, several 
transition metal oxides~\cite{Kargarian,Hsieh2}, Bi$_2$Te$_3$~\cite{Rauch}, and heavy-fermion 
compounds~\cite{Weng,Ye} have been found to be 3D TCIs.  

Rather recently, the TCI state has been theoretically predicted to occur in two dimensions (2D) 
in multilayer structures of SnTe~\cite{Liu}, PbSe~\cite{Wrasse}, and graphene 
multilayers~\cite{Kindermann}. While the experimental progress in 3D TCIs was 
rapid~\cite{Tanaka,Dziawa,Xu,Yan,Okada}, the 2D TCI phase disappears when placed on a substrate 
for further experimental investigation or device application due to the breaking of the mirror 
symmetry. Without the substrate, the freestanding film is usually hard to grow, which is also 
the main reason why 2D TIs were experimentally mainly established in HgTe/CdTe and InAs/GaSb 
quantum wells up to now~\cite{konig,Knez}. Therefore, it is highly desirable to search for 
new materials and novel formation mechanisms which maintain the mirror symmetry of 2D TCIs.

TCIs bear great potential for the investigation of exotic phenomena~\cite{Okada,Tang,Fang,fzhang}, 
such as the large Chern number quantum anomalous Hall (QAH) effect~\cite{Fang,fzhang}. 
The QAH state, which is characterized by a quantized charge conductivity without an external 
magnetic field, has topologically protected dissipationless chiral edge states and therefore 
is a good starting point to realise ultra-low-power consumption electronics~\cite{Haldane}. 
The TRS breaking in TIs provides a promising platform to investigate this striking topological 
phenomenon~\cite{yu,chang}.  However, the Chern number of the TI-based QAH state is usually 
limited to $\mathcal C= 1$~\cite{yu,chang, niu} or $\mathcal C = 2$~\cite{ZhangBi1,JWang}.  
In magnetically doped thin-film TCI SnTe, the QAH phase was predicted with the Chern number 
reaching as much as four~\cite{Fang,fzhang}. The QAH phase with large Chern number leads 
to multiple dissipationless edge channels that provide better ways to optimise electrical 
transport properties and could greatly reduce the contact resistance for circuit 
interconnects~\cite{Fang,JWang,Skirlo}. 

Here, based on density functional theory, we predict that SnTe and PbTe (001) monolayers are
2D TCIs. We expose our TCIs to an exchange field, which mimics the effect of 
magnetic adatoms or a substrate: While an in-plane exchange field destroys the TCI phase, 
it survives for an out of plane magnetization. Based on the considered materials, we 
demonstrate the feasibility of realizing a TCI-originated QAH phase with high Chern number, 
which is very robust to mirror symmetry breaking and/or magnetization fluctuations. Remarkably, 
a range of 2D TCIs can be realised in NaCl$_{2N}$(Sn/Pb)Te$_1$ quantum wells with a single 
layer of SnTe or PbTe embedded in $N$ layers of widely used substrate NaCl~\cite{Zemel} 
on both sides. These systems exhibit large topologically non-trivial energy gaps with a 
magnitude tunable by  the NaCl thickness. Our findings pave the way to utilization of 
TCIs as seed materials for arriving at robust topologically non-trivial phases.

The first-principles calculations have been performed using the full-potential linearized 
augmented-plane-wave method as implemented in the \texttt{FLEUR} code~\cite{fleur}. The 
self-consistent calculations with spin-orbit coupling (SOC) were carried out with a cutoff 
parameter, $k_{\rm max}$, of 3.8 bohr$^{-1}$. The experimental lattice parameters were 
used for pristine Sn/PbTe systems, while those for quantum well structures were obtained 
using the Vienna {\it ab-initio} simulation package (VASP)~\cite{Kresse}. The generalized 
gradient approximation in parametrization of Perdew, Burke and Ernzerhof (GGA-PBE) 
was used for the exchange correlation potential~\cite{Perdew}. The maximally localized 
Wannier functions (MLWFs) were constructed using the {\tt wannier90} code~\cite{Mostofi,Freimuth}.

\begin{figure}
\includegraphics{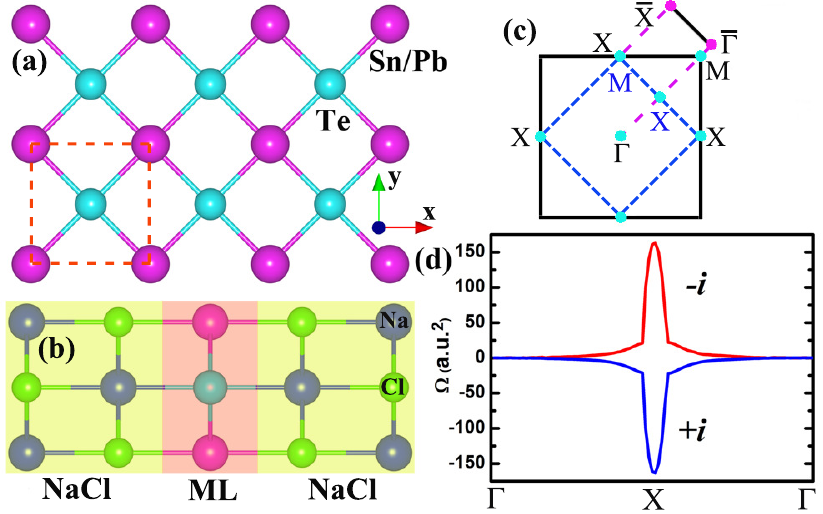}
\caption{(color online) 
 (a) Top view of the crystal structure of (001) Sn/PbTe monolayers with the unit cell indicated 
 by dashed lines. (b) Schematic view of the NaCl$_{2N}$(Sn/Pb)Te$_1$ quantum well structure for
  $N=2$. (c) Corresponding 2D and projected 1D Brillouin zones with marked high symmetry points. 
 (d) Berry curvature distribution associated with $\pm i$ mirror eigenstates of the occupied bands 
 along the $\Gamma-X-\Gamma$ path. }
\label{structure}
\end{figure}

Bulk SnTe and PbTe share the same face-centered cubic NaCl-type structure with an inverted 
band structure at the $L$ point, which results in the realisation of a 3D TCI state in bulk 
SnTe and a 2D TCI state in SnTe multilayers~\cite{Liu}. On the other hand, PbTe both in bulk 
and in thin (001) films is in the normal insulator (NI) state~\cite{Hsieh,Liu}. Here, we focus 
on the (001) oriented {\it monolayers} of SnTe and PbTe. In Fig.~\ref{structure}(a) the top 
view of such 2D monolayer is shown, with Sn (Pb) and Te atoms forming two square sublattices
positioned in the mirror plane $z=0$. Accordingly, all Bloch states in the system can be 
labeled with the eigenvalues $\pm i$ of the reflection operator with respect to this symmetry 
plane.

\begin{figure} [htbp]
\includegraphics{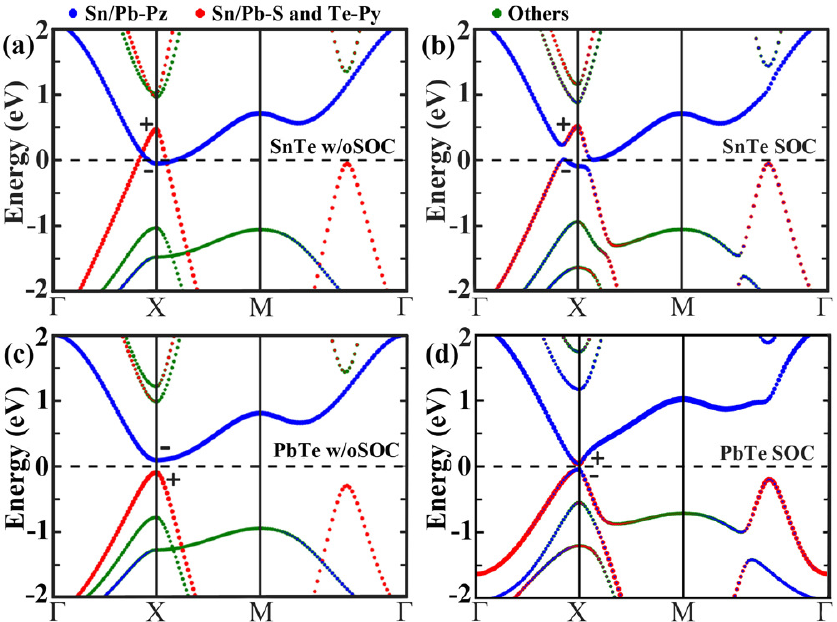}
\caption{(color online) 
 Orbitally-resolved band structures for (001) monolayers of SnTe (a, b) and PbTe (c, d) 
 without  (a, c) and with (b, d) SOC,  weighted with the contribution of $s$, $p_y$, and 
 $p_z$ states. Parities of the conduction band minimum and valence band maximum at the
 $X$ point are labeled by ``$+$" and ``$-$". The Fermi level is indicated with a dashed line.}
\label{band}
\end{figure}

To get a preliminary insight into the topological properties of the systems, we present 
in Fig.~\ref{band} the orbitally-resolved band structures of SnTe and PbTe monolayers with 
and without SOC. In the absence of SOC for SnTe, energy bands with the Sn-$s$ and Te-$p_y$ 
orbital character (positive parity with the inversion center at the Sn atom) overlap around 
the $X$-point with the Sn-$p_z$ states (negative parity). For PbTe at the $X$-point
without SOC a direct band gap appears with the valence band maximum (VBM) and the conduction 
band minimum (CBM) dominated by the Pb-$s$ and Te-$p_y$ orbitals (positive parity), and 
the Pb-$p_z$ orbital (negative parity), respectively. Turning on SOC leads to insulating 
character in both systems (calculated band gaps are 0.05 eV for SnTe and 0.09 eV for PbTe), 
and to the band inversion in PbTe, so that with SOC the band structure is inverted 
in both systems. 

Owing to an even number of $X$ points in the Brillouin zone (see Fig.~\ref{structure}(c)) 
neither the SnTe monolayer nor the PbTe monolayer is a 2D TI. However, taking the mirror 
symmetry into account, similarly to bulk  SnTe~\cite{Hsieh}, for both SnTe and PbTe 
monolayers band inversion results in the realisation of a 2D TCI state. To show this 
explicitly, we calculate the so-called mirror Chern number $n_{M} = (n_{+i}-n_{-i})/2$, 
where $n_{+i}$ and $n_{-i}$ are Chern numbers of all occupied bands with opposite mirror 
eigenvalues $+i$ and $-i$, respectively~\cite{Teo}. The Chern number of a given subset 
of states is given by ${\mathcal C}=\frac{1}{2\pi} \int_{\rm BZ} \Omega({\bf k})\, d^{2}k$, 
where $\Omega({\bf k})$ is the Berry curvature~\cite{Thouless,yao}:
\begin{eqnarray}
\Omega({\bf k})=\sum_{n<E_{F}}\sum_{m\ne n}2{\rm Im}\frac{\bra{\psi_{n{\bf k}}}\upsilon_x\ket{\psi_{m{\bf k}}}
       \bra{\psi_{m{\bf k}}}\upsilon_y\ket{\psi_{n{\bf k}}}}{(\varepsilon_{m{\bf k}}-\varepsilon_{n{\bf k}})^2},
\end{eqnarray}
with $m, n$ as band indices which run over considered subset of states, $\psi_{n{\bf k}}$ 
and $\varepsilon_{n{\bf k}}$ are corresponding Bloch states and their eigenenergies, and 
$\upsilon_{x/y}$ as the cartesian components of the velocity operator. In Fig.~\ref{structure}(d), 
we plot the distribution of the Berry curvature of all occupied bands with mirror eigenvalue $\pm i$. 
The main contribution to the Berry curvature comes from the region around $X$, with its values 
having opposite sign for opposite eigenvalues. The Chern number for each polarization is, 
respectively, $n_{+i}=-2$ and $n_{-i}=2$, yielding the total Chern number of all occupied 
states $\mathcal C = 0$  and the mirror Chern number $n_{M} = -2$. The calculated mirror 
Chern number $n_{M} = -2$ proves the TCI nature of (001) oriented SnTe and PbTe monolayers.
\footnote{The fact that in Ref.~\cite{Liu} only films thicker than four layers were found to 
be TCIs might be traced back to the fact that the film bandstructures were modeled from bulk
DFT results in that work. Here, DFT calculations were performed explicitly for the films.}

\begin{figure}
\includegraphics{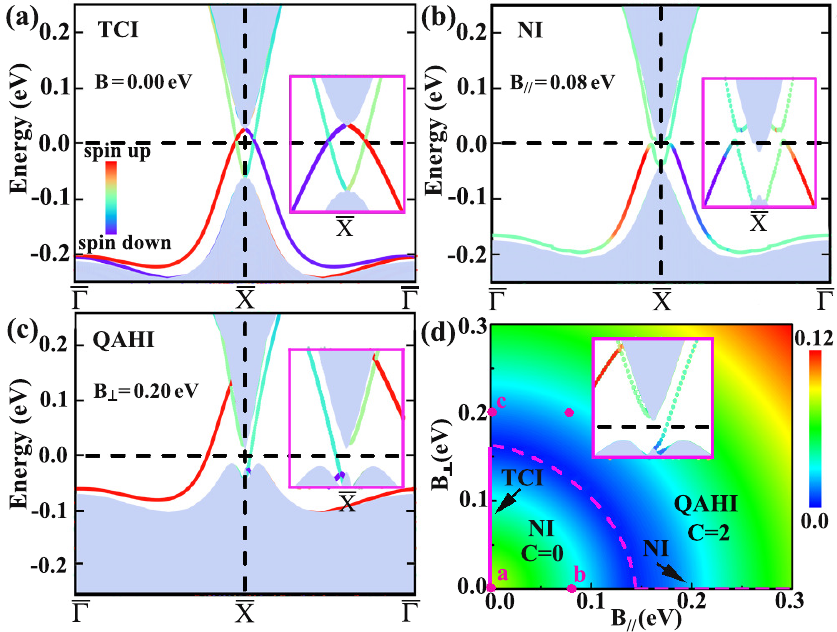}
\caption{(color online) 
Evolution of the edge states of a (110)-oriented ribbon of PbTe monolayer in an applied exchange 
field. (a) Upper-edge without exchange field ($B =0.00$~eV, 2D TCI), (b) Upper-edge with exchange 
field within the mirror plane ($B_{\parallel}=0.08$~eV, NI), and (c) upper-edge with exchange field 
perpendicular to the mirror plane ($B_{\perp}=0.20$~eV, QAHI). Edge states are colored with the 
expectation value of $\sigma_{z}$, as shown in (a). (d) Phase diagram of 2D PbTe with respect 
to $B_{\parallel}$ and $B_{\perp}$. The color scale represents the magnitude of the energy gap 
at the $X$ point (in eV). The dashed line marks the phase boundary and the dots correspond to 
situations shown in (a)-(c). The inset in (d) shows the chiral edge states at the upper-edge 
of the ribbon at a point in the phase diagram marked with a dot next to it.}
\label{edge}
\end{figure}

One of the prominent features of a TCI is the existence of gapless edge states that appear as a 
result of the crystal mirror symmetry only and do not require TRS~\cite{Fu,Hsieh}. To illustrate 
this, using description in terms of MLWFs of 2D monolayers, we construct the tight-binding 
Hamiltonian of finite 80-atom wide ribbons oriented along the (110)-axis. In addition, by 
computing the matrix elements of the Pauli matrices $\sigma_{\alpha} (\alpha = x, y, z)$ in 
the basis of MLWFs, we consider the effect of an exchange field applied to our ribbon, which 
breaks the TRS and generally has out of plane ($\sigma_{z}\cdot B_{\perp}$ term in the Hamiltonian) 
and in-plane ($\sigma_{x}\cdot B_{\parallel}$ term) components. By introducing such an exchange 
field we aim at mimicking the effect of interaction with a magnetic environment,~e.g.~magnetic 
adatoms or magnetic substrate~\cite{Inoue}.

The spin-resolved band dispersion of states localized at the ``upper" (``lower") edge and the 
projected bulk band structure without any exchange field is presented in (the inset of) 
Fig.~\ref{edge}(a). Clearly, according to the mirror Chern number analysis, two pairs of edge 
states cross slightly away from the time-reversal invariant $\bar X$-point, and the spin polarization
of the crossing states is of the same sign, which is in contrast to the case of 2D TIs.  For 
an in-plane exchange field the mirror symmetry is broken and the edge states become 
gapped (Fig.~\ref{edge} (b)). If the exchange field is, on the other hand, out of plane, the 
mirror symmetry survives and the system continues to exhibit gapless edge states, although the 
TRS is broken.

The TRS breaking causes an exchange splitting between conduction and valence bands of opposite 
spin, and thus, with increasing strength of the exchange field they approach each other. As 
shown in Fig.~\ref{edge}(d), with increasing the magnitude of $B_{\perp}$ the band gap decreases 
with the system remaining a TCI, and it closes at around $B_{\perp}=0.16$~eV, reopening again 
with further increasing $B_{\perp}$. According to our calculations, this signals a phase 
transition from a 2D TCI to a QAH phase, of which the latter is characterized by a non-zero 
Chern number of $\mathcal C=+2$. The nonzero Chern number is further verified by the presence 
of two gapless edge states on each side of the ribbon within the nontrivial band gap (see 
Fig.~\ref{edge}(c)). 

Our most interesting finding is that the QAH state in our TCI system can be reached irrespective 
of whether the mirror symmetry is broken by $B_{\parallel}$ or not (see computed phase diagram 
in Fig.~\ref{edge}(d) and the inset of it). This means that TCIs can be used to arrive at novel 
Chern insulators despite their apparent fragility to the mirror symmetry breaking,~e.g., by a  
substrate. In fact, it is only the magnitude of the exchange field and not its direction, which 
the evolution of the topological band gap is sensitive to, except for the situation with 
$B_{\perp}=0$, when the system always resides in the NI state. Such a peculiar behavior
also means that, when the QAH state is achieved in the system, it is not only robust with 
respect to the crystal symmetry, but also to the direction magnetization of adatoms or the
substrate, which is subject to thermal fluctuations.

\begin{figure}
\includegraphics{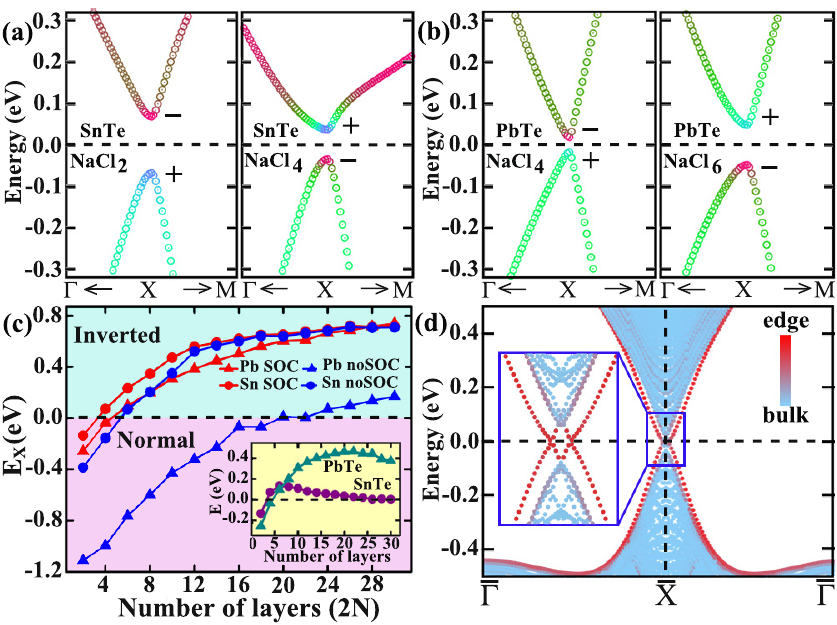}
\caption{(color online) 
Orbitally-resolved relativistic band structures of (a) NaCl$_{2N}$SnTe$_1$ and (b) NaCl$_{2N}$PbTe$_1$. 
Color from red to blue represents the increasing contributions of Sn/Pb-s and Te-$p_y$ orbitals, which 
reveals the band inversion clearly. (c) Calculated energy gap of NaCl$_{2N}$(Sn/Pb)Te$_1$  superlattice at 
the $X$ point with and without SOC as a function of NaCl thickness. Negative and positive values 
correspond to the normal and inverted energy gaps, respectively. Inset shows the global energy gap as 
a function of NaCl thickness. (d) Localization-resolved band structure of a 1D nanoribbon of 
NaCl$_6$PbTe$_1$. Inset is the zoom-in at the $\overline{X}$ point.}
\label{nacl}
\end{figure}

Finally, we choose the widely used material NaCl~\cite{Zemel} to propose a practical way of stabilizing 
the 2D TCIs while keeping the mirror symmetry in NaCl$_{2N}$(Sn/Pb)Te$_1$ quantum wells. One of our
main findings here is that the magnitude of the TCI band gap can be engineered by the thickness of 
NaCl spacer, and it can reach as much as 0.13 eV and 0.47 eV in NaCl$_{2N}$SnTe$_1$ and NaCl$_{2N}$PbTe$_1$ 
wells, respectively. The sketch of the crystal structure of a NaCl$_{2N}$(Sn/Pb)Te$_1$  quantum well 
is shown in Fig.~\ref{structure} (b). A SnTe (PbTe) monolayer is placed in between  $N$-layers thick 
NaCl films, so that the mirror symmetry survives. When SnTe (PbTe) monolayer is sandwiched between 
NaCl slabs, the interaction between SnTe (PbTe) and NaCl pushes the Sn/Pb-$p_z$ orbital up in energy 
and Sn/Pb-$s$ with Te-$p_y$ orbitals down in energy at the $X$ point. For the case of $N=1$ the 
strongest interaction is observed, resulting in the disappearance of the band inversion and the TCI 
phase. With increasing thickness of NaCl layers, the strength of the local  potential decreases. 
Consequently, the band inversion and the TCI phase return at a critical NaCl thickness. 

To show the thickness-dependent band inversion, we plot the orbitally-resolved band structures in 
Fig.~\ref{nacl}(a,b). We observe that for small $N$ the Sn/Pb-$s$ and Te-$p_y$ orbitals contribute 
to the VBM with positive parity, while, at larger NaCl thickness, they contribute to the CBM keeping
positive parity.  Therefore, the band inversion occurs at $N=2$ and $N=3$ for the NaCl$_{2N}$SnTe$_1$
and NaCl$_{2N}$PbTe$_1$, respectively. Fig.~\ref{nacl}(c) shows the magnitude of the energy gaps at 
the $X$ point without and with SOC versus the NaCl thickness. For NaCl$_{2N}$SnTe$_1$, the existence 
of the normal energy gap in both cases demonstrates a normal insulator phase for $N=1$. For $N=2$, a 
SOC-induced band inversion appears, indicating the realisation of the 2D TCI phase. As the NaCl 
thickness increases further, the energy gap at the $X$ point increases and the inverted energy gap is 
obtained even without SOC.  Similar behaviour is seen for NaCl$_{2N}$PbTe$_1$ with varying $N$, but 
the influence of SOC is more pronounced in this case. The 2D TCI state is further explicitly confirmed 
by the topological analysis and the emergence of the gapless edge states in 1D nanoribbons (see 
Fig.~\ref{nacl}(d)). The global energy gap is one of the defining properties of TCI. As shown in 
the inset of Fig.~\ref{nacl}(c), the global gap survives for a large range of NaCl thickness, and 
its magnitude can be tuned by a proper choice of $N$, reaching as much as 0.13 eV and 0.47 eV for 
NaCl$_6$SnTe$_1$ and NaCl$_{22}$PbTe$_1$, respectively. We note that the latter value
of 0.47 eV is the largest for either 2D or 3D TCIs reported so far. 

In summary, we have identified by first-principles calculations that the (001) oriented SnTe and PbTe 
monolayers are topological crystalline insulators. We explore the phase diagram of these materials with 
respect to an applied exchange field due to magnetic adatoms/substrate. We show that as the mirror 
symmetry remains for the out of plane magnetization, the gapless edge states survive, but the TCI state 
is destroyed for an in-plane exchange field. For finite $B_{\perp}$, upon reaching a certain strength 
of the exchange field we show the emergence of a robust Chern insulator phase which survives 
irrespective of the mirror symmetry breaking. We further show that the large-gap TCI Pb/SnTe
based family can be successfully stabilised in NaCl$_{2N}$(Sn/Pb)Te$_1$ quantum wells.  Especially 
for NaCl$_{22}$PbTe$_1$, an energy gap of 0.47 eV appears, which is by far larger than that in 
known 2D and 3D TCIs. 

\acknowledgments{We are grateful to Hongbin Zhang for insightful discussions. This work was supported 
 by the SPP 1666 of the German Research foundation (DFG) and the HGF-YIG Programme VH-NG-513. 
 We acknowledge computing time on the supercomputers JUQUEEN and JUROPA at J\"{u}lich Supercomputing 
 Centre and JARA-HPC of RWTH Aachen University.}

\end{document}